\documentclass[aps,prd,showpacs]{revtex4}

\usepackage{amsmath,amsfonts,amssymb}
\usepackage[dvips]{graphicx}

\newcommand{\ud}{\mathrm{d}}

\begin{document}

\title{Dynamical dark energy models -- dynamical system approach}
\author{Marek Szyd{\l}owski}
\email{uoszydlo@cyf-kr.edu.pl}
\affiliation{Complex Systems Research Center, Jagiellonian University,
Reymonta 4, 30-059 Krak{\'o}w, Poland}
\author{Orest Hrycyna}
\email{hrycyna@byk.oa.uj.edu.pl}
\affiliation{Astronomical Observatory, Jagiellonian University,  
Orla 171, 30-244 Krak{\'o}w, Poland}

\begin{abstract}
We study the Friedmann-Robertson-Walker model with dynamical dark energy modelled in terms of the equation of state $p_{x}=w_{x}(a(z)) \rho_{x}$ in which the coefficient $w_{x}$ is parameterized by the scale factor $a$ or redshift $z$. We use methods of qualitative analysis of differential equation to investigate the space of all admissible solutions for all initial conditions on the two-dimensional phase plane. We show advantages of representing this dynamics as a motion of a particle in the one-dimensional potential $V(a)$. One of the features of this reduction is the possibility of investigating how typical big rip singularities are in the future evolution of the model. The properties of potential function $V$ can serve as a tool for qualitative classification of all evolution paths. Some important features like resolution of the acceleration problem can be simply visualized as domains on the phase plane. Then one is able to see how large is the class of solutions (labelled by the inset of the initial conditions) leading to the desired property.
\end{abstract}

\pacs{}

\maketitle

\section{Introduction}

While the cosmic acceleration becomes one of the most exciting discoveries of the cosmology \cite{Riess:1998,Perlmutter:1999} the nature of the dark energy which is still unknown \cite{Ratra:1988,Caldwell:1998}. Unfortunately, there are very weak constraints on its form of an equation of state \cite{Perlmutter:1999,Bean:2002}. The most popular candidate for dark energy is the cosmological constant $\Lambda$, which can be treated as some kind of perfect fluid satisfying the equation of state $p=-\rho$, $\rho=\Lambda$. However, it remains to explain why value of $\Lambda$, obtained from type Ia supernovae (SNIa) observations, is so small in comparison with the value of $\Lambda$ interpreted as vacuum energy (i.e., the Planck mass scale). Recent investigations in this field concentrate mainly on the dark energy, modelled as $(a)$ quintessence scalar field \cite{Peebles:2003} or $(b)$ based on barotropic equation of state \cite{Kamenshchik:2000}. In this paper we assume the general form of the equation of state in which the equation of state factor $w$ of total matter depends on redshift $z$ through the scale factor $a$, so $w(a(z)) = p/\rho$, $w < -1/3$ in some nonempty interval of evolution.

It is characteristic the existence of large variety of different dark energy models \cite{Peebles:2003}. We unify all these models assuming that dark energy obeys the general form of the equation of state $p=w(a(z)) \rho$. It is interesting to find some general properties of an evolutional path of such models. For this aim dynamical system methods seem to be a natural method because it offers the possibility of investigating whole space of solutions starting from all admissible initial conditions. The philosophy of qualitative investigation of differential equations shifts interest from founding their exact solutions toward investigating their sensitivity (or fragility) with respect to small changes of initial conditions or model parameters.

It is demonstrated that dynamics of the FRW model with dark energy can be represented in the form of two-dimensional dynamical system, $\dot{x}=P(x,y)$, $\dot{y}=Q(x,y)$ where $P,Q \in C^{\infty}$. The phase portraits of this system are organized by critical points $(x_{0},y_{0}) \colon P(x_{0},y_{0})=Q(x_{0},y_{0})=0$ or limit cycles (non-point-like attractors). Following the Hartman-Grobman theorem in neighborhood of non-hyperbolic critical points \cite{Perko:1991} ($ Re \lambda_{i} \ne 0$, where $\lambda_{i}$ are eigenvalues of linearization matrix $\dot{x}=A x$, $A$ is the Jacobi matrix) the nonlinear matrix is equivalent to its linearization. The full knowledge of dynamical behavior requires analyzing its behavior at infinity. To achieve this it is useful to transform trajectories from a $\mathbf{R}^{2}$ phase plane into a Poincar{\'e} sphere \cite{Perko:1991}. Then infinitely distant points of the plane are mapped into a sphere's equator $S^{1}$. The type of critical points are conserved under this mapping but critical points present in infinity can appear at the equator. Hence an orthogonal projection of any hemisphere onto the tangent plane gives a compactified phase portrait.

The main subject of this paper is presentation how dynamics of the FRW model with dynamical dark energy can be reduced to the form of a two-dimensional dynamical system (section~\ref{sec:2}). The general properties of such a system are investigated in section~\ref{sec:3}. We find that a global structure of the compactified phase space $\mathbf{R}P^{2}$ depends on the strong energy condition only. In section~\ref{sec:4} we present advantages of using a complementary description of dynamics as a Hamiltonian flow. In this picture the evolution of the model is represented as a motion of unit mass in the one-dimensional potential $V(a)$. It is interesting that a shape of the diagram of $V(a)$ contain all informations which are needed to determine all critical points at the finite domain as well as at infinity and their character. The detailed discussion of this issue and the compactified phase portraits contain section~\ref{sec:5}. We find only three generic cases of the phase portraits. All FRW models with a postulated form of dark energy belong to the one from distinguished class of phase portraits modulo homeomorphism preserving orientation of the phase curves (which establish equivalence of the phase portraits). Therefore it is possible to classify all FRW dark energy models on the phase portraits without some details about the specific form of the equation of state.

\section{The FRW models with dynamical dark energy as a dynamical systems}
\label{sec:2}

Let us consider the FRW model with source in the form of noninteracting dust matter and dark energy which the equation of state is parameterized by $p_{x}=w_{x}(a) \rho_{x}$, $w_{x} < -1/3$. Then from the conservation condition
\begin{equation}
\dot{\rho}=-3 \frac{\dot{a}}{a}(\rho + p),
\label{eq:1}
\end{equation}
where a dot denotes the differentiation with respect to time $t$, we obtain for each component relations
\begin{equation}
\begin{array}{ll}
\rho_{m}=\rho_{m,0} a^{-3}, & \rho_{m}(a=1)=\rho_{m,0}; \\
\rho_{x}=\rho_{x,0} a^{-3} \exp{\Big(-3 \int_{1}^{a} \frac{w_{x}(a)}{a} \ud a\Big)}, & \rho_{x}(a=1)=\rho_{x,0}.
\end{array}
\label{eq:2}
\end{equation}
Hence we can combine both effects and define the total pressure as 
\begin{equation}
p = 0 + w_{x} \rho_{x} \equiv w(a)\rho,
\label{eq:3}
\end{equation}
where the equation of state factor for total matter is
\begin{equation}
w(a) = \frac{w_{x}(a)}{1+\displaystyle{\frac{\Omega_{m,0}}{\Omega_{x,0}}} \exp{\Big(3 \int_{1}^{a} \frac{w_{x}(a)}{a} \ud a \Big)}},
\label{eq:4}
\end{equation}
where $\Omega_{i,0}$ are density parameters for both matter and dark energy, $a$ is expressed in the units of its present value $a_{0}$.

In our further analysis we assume the equation of state in form (\ref{eq:3}). Note that the coefficient of state $w_{x}(a)$ can be always extracted from the integral differential equation
\begin{equation}
w_{x}(a) = w(a) \bigg\{ 1+\frac{\Omega_{m,0}}{\Omega_{x,0}} \exp{\Big(3 \int_{1}^{a} \frac{w_{x}(a)}{a} \ud a \Big)} \bigg\},
\label{eq:5}
\end{equation}
equivalent to the differential equation
\begin{equation}
\frac{\ud y}{\ud a} = 3 y (y - 1) \frac{w(a)}{a}, \qquad y=\frac{w_{x}(a)}{w(a)},
\nonumber
\end{equation}
which the solution, for the initial condition $w(1)/w_{x}(1)=\Omega_{x,0}/(\Omega_{x,0}+\Omega_{m,0})$, is in the form
\begin{equation}
y \equiv \frac{w_{x}(a)}{w(a)} = 1+\frac{\Omega_{m,0}}{\Omega_{x,0}} \exp{\Big(3 \int_{1}^{a} \frac{w_{x}(a)}{a} \ud a -1\Big)}.
\label{eq:6}
\end{equation}
Due to (\ref{eq:6}) it is possible to determine $w_{x}(a)$ provided that $w(a)$ is given.

The FRW dynamics is governed by equations
\begin{equation}
\begin{array}{l}
\displaystyle{\rho=3 \frac{k}{a^{2}} +3 \frac{\dot{a}^{2}}{a^{2}} ,} \\
\displaystyle{p = -2 \frac{\ddot{a}}{a} - \frac{\dot{a}^{2}}{a^{2}} - \frac{k}{a^{2}} ,}
\end{array}
\label{eq:7}
\end{equation}
where $\rho=\rho_{\textrm{eff}}$ and $p$ are effective energy density and pressure respectively; $\rho_{\textrm{eff}} = \rho_{m}+\rho_{x}$ in the model under consideration, $k$ -- the curvature index $k=0,\pm 1$. System (\ref{eq:7}) together with condition (\ref{eq:1}) is closed if the form of the equation of state (\ref{eq:3}) is postulated.

Then the basic dynamical equations reduce to 
\begin{equation}
\begin{array}{l}
\displaystyle{\dot{a} = y }, \\
\displaystyle{\dot{y} = - \frac{1}{6} \{ \rho(a) + 3 p(a) \} a}.
\end{array}
\label{eq:8}
\end{equation}
Note that expression
\begin{equation}
\frac{y^{2}}{2} + V(a) = - \frac{k}{2},
\label{eq:9}
\end{equation}
plays the role of first integral (\ref{eq:8}) if
\begin{equation}
V(a) = - \rho_{\textrm{eff}} \frac{a^2}{6}.
\label{eq:10}
\end{equation}
Equations (\ref{eq:8}) constitute a two-dimensional autonomous dynamical system which solutions can be visualized in the phase plane $(a,y)$. Then (\ref{eq:9}) represents the equation for an algebraic curve on which lies solutions of the dynamical system. Among all solutions there exist singular solutions represented by critical points in the phase plane. At the finite domain of the phase plane they always represent static universes $(\dot{a}=0)$
\begin{equation}
y_{0}=0 \quad \textrm{and} \quad \rho(a_{0})+ 3 p(a_{0})=0.
\label{eq:11}
\end{equation}
They are located on $a$-axis, and exist if only $p < -\rho/3$ for effective pressure and energy density. There are two possibilities: $\rho(a_{0}) = 0$ or $w(a_{0}) = -1/3$ for appearing such a points. If $w(a)$ is a monotonic function of $a$ then there is one solution $a_{0}\colon w(a_{0}) = -1/3$. Note that for all critical points the function $V(a_{0}) = -k/2$. Because it is always the non-positive function if only $\rho_{\textrm{eff}} \ge 0$ the critical points are admissible in the region occupied by trajectories of the closed model $(k=+1)$.

Following the first integral (\ref{eq:9}) the whole phase space $(a,y)$ is divided on two disjoint domains by the trajectory of the flat model. In these regions we find closed models in the region $y^{2}/2 + V(a) \le 0$ while open models are situated in the region $y^{2}/2 + V(a) \ge 0$ of the phase space. The boundary of the strong energy condition $\rho + 3 p =0$ defines the line in the phase space which separates regions occupied by non-accelerating and accelerating phases of evolution of the models.

\section{The general properties of the FRW models with dynamical dark energy}
\label{sec:3}

The assumed form of the equation of state includes different dark energy models. For example the Chaplygin gas can be treated as a special case of some prescribed form of the equation of state \cite{Kamenshchik:2000}. In this case we have
\begin{equation}
p = - \frac{A}{\rho} = w(a) \rho,
\label{eq:12}
\end{equation}
where $w(a) = - A a^{6}/(A a^{6} + B)$ and $A$, $B$ are positive constants. From (\ref{eq:12}) one can seen that $w$ varies in the interval $[-1,0]$. It is a  monotonic (decreasing) function of $a$.

If we consider the equation of state for a non-interacting mixture of dust ($p=0$) and radiation matter then we obtain the equation of state in the form
\begin{equation}
p = 0 + \frac{1}{3} \rho_{r} = \frac{1}{3} \frac{\rho_{r}}{\rho_{r}+\rho_{m}} \rho = \frac{1}{3} \frac{1}{1+ \displaystyle{\frac{\Omega_{m,0}}{\Omega_{r,0}} a}} \rho ,
\label{eq:13}
\end{equation}
which can be reproduced by $w=w(a)$
\begin{equation}
w(a)=\frac{1}{3} \frac{1}{1 + \displaystyle{\frac{\Omega_{m,0}}{\Omega_{r,0}} a}}.
\label{eq:14}
\end{equation}
Freese and Lewis \cite{Freese:2002} have recently considered models (called Cardassian) which are alternative to dark energy models. Following their approach the source of acceleration is in some modification of FRW dynamics instead of dark energy conditions. One can show that the Cardassian models are equivalent to dark energy models for two components of matter. The basic equations are
\begin{equation}
\begin{array}{l}
\rho_{\textrm{eff}} = (\rho_{m} + \rho_{r}) + B (\rho_{m} + \rho_{r})^{n}, \\
3 H^{2} = \rho_{\text{eff}},
\end{array}
\label{eq:15}
\end{equation}
where $H=(\ln{a})\dot{}$ is the Hubble function and $\rho_{r}$ is energy density of radiation matter: $k=0$ is the assumption of the model, $n$ is the model parameter and
\begin{equation}
p_{\textrm{eff}} = w(a)(\rho_{m} + \rho_{r}) + \bar{w}(a)(\rho_{m} + \rho_{r})^{n},
\label{eq:16}
\end{equation}
where $\bar{w}(a) = n (w + 1) - 1$ and $w(a)$ is given by formula (\ref{eq:14}). Therefore, assumed for (\ref{eq:3}) of equation of state seems to be sufficiently general to unify all cosmological models which offers explanation of SNIa data (note that it is also valid of the Linder and Copeland model).

Because we assume the presence of dust matter we have $x_{0} \ne 0$ but in general for the fluid violating the weak energy condition $(\rho + p)(x) > 0$ it is possible to have the critical point $x_{0}=y_{0}=0$. Then $\det{A} = 1/6(\rho + 3 p)|_{(0,0)}$ and $\lambda^{2} + \det{A}=0$, therefore eigenvalues of the linearization matrix are real of positive signs which correspond to saddles points. Note that in this case there are no static critical points at $\rho + 3p = 0$ in the finite domain.

The phase space of the model is organized by critical points and phase curves determined from the first integral (\ref{eq:9}). The critical points at the finite domain of the phase space are situated on the $a$-axis as intersection points of the $a$-axis with the boundary curve of the strong energy condition  $\rho + 3 p \ge 0$. The character (type) of critical points is determined from eigenvalues of the linearization matrix $A$:
\begin{equation}
A = \left[ \begin{array}{cc}
\displaystyle{\frac{\partial P}{\partial x}} & \displaystyle{\frac{\partial P}{\partial y}} \\[9pt]
\displaystyle{\frac{\partial Q}{\partial x}} & \displaystyle{\frac{\partial Q}{\partial y}}
\end{array} \right]_{(x_{0},y_{0})}
= \left[ \begin{array}{cc}
0 & 1 \\
\displaystyle{-\frac{1}{6} \frac{\partial}{\partial x} \big(x \rho(x)(1 + 3 w(x))\big)} & 0 
\end{array} \right]_{(x_{0},y_{0})} ,
\label{eq:17}
\end{equation}
where $x=a$ is a dimensionless scale factor expressed in terms of its present value $a_{0}$.

The eigenproblem $\det{[A - \lambda \mathbf{1}]}=0$ then reads $\lambda^{2} - \lambda \textrm{Tr}A + \det{A} = 0$. Consequently the sign of the determinant $A$ determines the type of the critical points, i.e., whether $\lambda$ is real or complex. It is consequence of the fact that 
\begin{equation}
\begin{array}{l}
\textrm{Tr}A = 0 \quad \textrm{and} \quad \det{A}= \frac{1}{6}\frac{\partial}{\partial x} \big(x \rho(x)(1 + 3 w(x))\big) \\
\lambda^{2} + \det{A} =0.
\end{array}
\label{eq:18}
\end{equation}
Because at the critical points we have $(\rho + 3p)(x_{0})=0$ we have
\begin{equation}
\det{A} = x_{0} \frac{\partial}{\partial x}\bigg|_{x_{0}}(\rho + 3p)(x).
\label{eq:19}
\end{equation}
The critical points of the system can be saddle points if $\det{A}|_{x_{0}} < 0$ or centers in the opposite case if $\det{A}|_{x_{0}} > 0$. In the first case eigenvalues are real of opposite signs while in the second case they are purely imaginary and conjugated. The determinant of the linearization matrix can by rewritten to the form
\begin{equation}
\det{A} = \frac{\partial \rho(x)}{\partial x}\bigg|_{x_{0}}(1 + 3w(x))\bigg|_{x_{0}} + 3 \rho(x_{0}) \frac{\ud w}{\ud x}\bigg|_{x_{0}}.
\label{eq:20}
\end{equation}
Therefore if we consider critical points appearing at $w(x_{0})=-1/3$ then $\det{A}=3 \rho(x_{0})\ud w/\ud x |_{x_{0}}$ and the type of critical points depends on sign of $\ud w/\ud x$ at the critical point $x_{0}$ (we assume $\rho(x) > 0$). If $w(x)$ is a decreasing function of the scale factor then the critical points are saddle types.

If we consider the second type of critical points which is the static critical point located at $x=x_{0}$ such that $\rho(x_{0})=0$ then first term in (\ref{eq:20}) decides whether the critical points are saddles or centers. Therefore if $x_{0} \colon w(x_{0}) < -1/3$ ($x_{0} \colon w(x_{0}) > -1/3$) then $\det{A} < 0$ ($\det{A} > 0$) if $\ud \rho/\ud x |_{x_{0}} > 0$ (like for phantom fields violating the weak energy condition $\rho + p > 0$). If $x_{0}$ is located in the decelerating region $x_{0} \colon w(x_{0}) > -1/3$ then $\det{A} < 0$ ($\det{A} > 0$) if $\ud \rho/\ud x |_{x_{0}} < 0$ (like for matter satisfying a weak energy condition). If $\det{A} > 0$ or $\det{A < 0}$, centers or saddles are admissible respectively. Note that both energy density of matter and dark energy are positive this case is excluded at finite domains. Finally one can conclude that if we assume positivity of energy of each component then the existence as well as character of the critical points are determined by energy conditions in the neighborhood of the transition moment corresponding the inflection of the diagram $a(t)$. From the physical point of view it is moment of hanging decelerating phase into accelerating one.

The full knowledge of dynamical behavior requires an analysis of the system at infinity. One can perform such an analysis in a simple way by introducing the projective coordinates on the phase plane. There are two maps which cover a circle $S^{1}$ at infinity
\begin{equation}
\begin{array}{cl}
\textrm{(1)} & \displaystyle{(z,u) \colon \quad z = \frac{1}{x} , \quad u = \frac{y}{x} , \quad z = 0 , \quad -\infty < u < +\infty}, \\[9pt]
\textrm{(2)} & \displaystyle{(v,w) \colon \quad v = \frac{1}{y} , \quad w = \frac{x}{y} , \quad v = 0 , \quad -\infty < w < +\infty}.
\end{array}
\label{eq:21}
\end{equation}
By adding a circle $S^{1}$ at infinity to $\mathbf{R}^2$ we obtain a compact space -- projective plane $\mathbf{R}P^{2}$. In the $(z,u)$ coordinates the system under consideration has the following form
\begin{equation}
\begin{array}{l}
\displaystyle{\frac{\ud z}{\ud t} = -z u} ,  \\[9pt]
\displaystyle{\frac{\ud u}{\ud t} = -\frac{1}{6}(\rho + 3 p) - u^{2}},
\end{array}
\label{eq:22}
\end{equation}
with the first integral in the form $u^{2}+ 2 z^{2} V(1/z) = k z^{2}$. There are two types of critical points situated on the circle at infinity ($z=0$,$-\infty < v < +\infty$)
\begin{equation}
z_{0}=0, \quad u_{0}=\pm \sqrt{-\frac{1}{6}(\rho + 3p)_{0}}.
\label{eq:23}
\end{equation}
Of course such points are admissible if the strong energy condition at the critical point is violated $(\rho + 3 p)_{0} < 0$ and the value of $(\rho + 3 p)_{0}$ is finite (we can also find a critical point at a finite domain which was detected previously $u_{0} = 0$ ($H=\infty$) and $(\rho + 3p)_{0} = 0$). From the form of the first integral of (\ref{eq:22}) we obtain that critical points (\ref{eq:23}) are intersection points of the trajectory of the flat model $u^{2} + 2 z^{2} V(1/z) = 0$ with the circle at infinity. They are representing stationary solutions $H_{0} = \pm \sqrt{-1/6(\rho+ 3p)_{0}}$, i.e., the expanding and contracting deSitter solutions. For existence of such points it is required that $z^{2} V(1/z)$ goes to a constant value as $z \to 0$. For example if we consider the $\Lambda$CDM model with the cosmological term $\Lambda$ then $\rho(z=0)=\Lambda$, $H_{0}=\pm \sqrt{1/3}$. In the special case if $z^{2} V(1/z) \to 0$ as $z \to 0$ we obtain a static solution.

For the analysis of the type of critical points we consider the linearization matrix 
\begin{equation}
A = \left( \begin{array}{cc}
-u_{0} & 0 \\
-\frac{1}{6} \frac{\ud}{\ud z}(\rho + 3 p)\big|_{(0,u_{0})} & -2u_{0}
\end{array} \right)
\label{eq:24}
\end{equation}
In this case $\textrm{Tr}A=-3u_{0}$ which means that if $u_{0}>0$ (expanding deS$_{+}$) the deSitter space is a global attractor, while if $u_{0} < 0$ deS$_{-}$ is representing a global repellor. If $u_{0} = 0$ all eigenvalues $\lambda_{1,2} = 1/2 (-3 u_{0} \pm |u_{0}|)$ are degenerated, non-hyperbolic critical point.

Analogical investigations of the critical points at infinity can be performed in the map $(v,w) \colon v=1/y, w=x/y$. The dynamical system (\ref{eq:8}) in these coordinates is of the form
\begin{equation}
\begin{array}{l}
\displaystyle{\frac{\ud v}{\ud t} = \frac{v w}{6}(\rho + 3p)(\frac{w}{v})}, \\[9pt]
\displaystyle{\frac{\ud w}{\ud t} = 1 + \frac{w^{2}}{6}(\rho + 3p)(\frac{w}{v})},
\end{array}
\label{eq:25}
\end{equation}
with first integral $1/(2 v^{2}) + V(w/v) = -k/2$.

At circle at infinity $v=0$ the effects of curvature are negligible. Therefore the critical points at infinity are intersections points of the algebraic curve $1 + 2 v^{2} V(w/v) = 0$ with a circle at infinity $v=0$ $(y=\infty)$. If the function $V$ is negative and for large $x$ behaves asymptotically like $V(x) \propto x^{m}$, i.e., it is a homogeneous function of degree $m$, then for $m=0$ deS attractors (repellors) are admissible. If $m>2$ then there are two critical points in the domain $y>0$ and their counterparts in the symmetric domain $y<0$ (with respect to an $x$-axis). They lie on the trajectory of the flat model and a circle $S^{1}$. For both critical points the projective coordinates are the same $v_{0}=0$, $w_{0}=0$. If we consider $V(x)$ in the neighborhood of $x=0$ in a power law approximation: $V(x) \propto x^{m}$, $m<0$ then the critical point $x=0$, $y=\infty$ $(v=0)$ can be recovered. In the neighborhood of this critical point, representing an initial singularity, dust matter effects dominate while the effects of dark energy (for which $w_{x} < -1/3$) are negligible. Big rip singularities arise in the phantom dark energy cosmological model \cite{Caldwell:2003}. At a big rip type of singularities the cosmological scale factor and the Hubble function achieve an infinite value in a finite interval of time. For both critical points at infinity $y=\pm \infty$ (also $H=\infty$) but for the big-rip singularity (in contrast to big bang when $x=0$) we have $x=\infty$ (or $w=0$). If we put $m=2+\varepsilon$, where $\varepsilon > 0$ then $y \propto x^{1+\varepsilon}$ approximate trajectories near big-rip singularity while $y/x$ achieve infinite value at big-rip singularity as well as at the initial singularity the values of $x$ are infinity and zero respectively, so they appear as a different point on the phase plane $(x,y)$.

Finally one can conclude that existence and type of critical points at the finite domain an at infinity is determined by the strong energy conditions (at finite domain) and the trajectory of the flat FRW model. The intersection trajectory of the flat model with a circle at infinity determines the location of the critical point at infinity. We assume a two-components model of matter filling the Universe. In the neighborhood of the initial singularity $x=0$ dust matter dominate the effects of dark energy. Therefore for the late time dark energy ($w_{x} < -1/3$) should be feasted. Three different scenarios are possible if dynamical behavior for late time is admissible. It depends whether $w_{x} + 1 \gtrless 0$ and consequently the sign of $w_{x} + 1$ determines the type of a global attractor (a deSitter is $w_{x}=-1$ or a big-rip attractor if $w_{x} < -1$ or a model with asymptotic $H=0$). From the form of first integral (\ref{eq:9}) one can observe symmetry of the phase space $y \to -y$. For the case of $w_{x} > -1$ there is no deSitter attractor but trajectories going toward the state $a_{0} \colon \rho(x_{0})$---the static critical point at infinity. Note that while the localization of the static critical point and its character depends on solutions of the equation $(\rho + 3p)(a)=0$ and $ \ud/\ud a |_{a=0} (\rho + 3p)$ in the neighborhood of $a_{0}$, the behavior at infinity ($t \to \infty$) depends crucially on parameter $m$ in asymptotic of $V \propto x^{m}$.

Note that this type of the future singularity may appear for the pressure ($w=-\infty$) and not only for the energy density $\rho$ (see also \cite{Barrow:2004}). Note that because our system is autonomous it has symmetry $t \to t + b$ , $b=$const it is always possible that a singularity occurs for some finite value of the cosmological time. In other words any two phase portraits with rescaled time (through a smooth function) are the same.

\section{Accelerating FRW Universe as a particle in a one-dimenensional potential}
\label{sec:4}

In this section we adopt the Hamiltonian formalism to our dynamical problem. This gives at once the insight into the previously introduced function $V(x)$. Moreover our problem stays similar to that of a particle moving in the one-dimensional potential energy $V(x)$. To prove this let us rewrite Eq. (\ref{eq:8}) to the form analogous to the Newtonian equation of motion
\begin{equation}
\ddot{x} = -\bigg(\frac{\partial V}{\partial x}\bigg), \quad V(x) = \frac{1}{6}\int_{0}^{x}(\rho + 3p)a \ud a.
\label{eq:26}
\end{equation}

If we assume that $\rho(a)$ and $p(a)$ satisfy the conservation condition (\ref{eq:1}) the integration by part can be exactly performed. Then, after including (\ref{eq:1}) rewritten to the equivalent form
\begin{equation}
\frac{\ud}{\ud t}\big(\rho a^{3}\big) + p \frac{\ud}{\ud t}\big(a^{3}\big) = 0,
\label{eq:27}
\end{equation}
which enables us to eliminate $p$ from $V$ given by (\ref{eq:26}), we obtain, as a result of integration,
\begin{equation}
V(a) = -\rho \frac{a^{2}}{6},
\label{eq:28}
\end{equation}
where if $a$ is expressed in the units $a_{0}$, $V(a)=V(x)$.

Of course, system (\ref{eq:26}) has the first integral
\begin{equation}
\frac{\dot{x}^{2}}{2} + V(x) = \frac{1}{2} \Omega_{k,0},
\label{eq:29}
\end{equation}
where here a dot denotes differentiation with respect to rescaled $t$ time $\tau \colon t \to \tau \colon \ud t = |H_{0}|^{-1} \ud \tau$, $\Omega_{k,0} = - k/a_{0}^{2}$ is the density parameter for curvature fluid, $V(x) = -(1/2) \Omega_{m,0} x^{-1} - \Omega_{x} x^{2}$;  $\Omega_{m,0}$ and $\Omega_{x}$ are density parameters for matter and dark energy, respectively.

Therefore, from Eq. (\ref{eq:29}) we find interpretation of $V(x)$ as a potential function for a particle-universe moving in the configuration space ${x \colon x \ge 0}$. Equation (\ref{eq:26}) can be simply reduced to the two-dimensional dynamical system
\begin{equation}
\begin{array}{l}
\displaystyle{\dot{x} = y}, \\
\displaystyle{\dot{y} = - \frac{\partial V}{\partial x}} , \\
\displaystyle{\frac{y^{2}}{2} + V(x) = \frac{1}{2} \Omega_{k,0}}.
\end{array}
\label{eq:30}
\end{equation}

In terms of the potential critical points are represented as extremal points: $\partial V / \partial x |_{x_{0}} = 0$. The linearization matrix and eigenvalues are
\begin{equation}
A = \left( \begin{array}{cc}
0 & 1 \\
\displaystyle{-\frac{\partial^{2} V}{\partial x^{2}}} & 0
\end{array} \right),
\label{eq:31}
\end{equation}

\begin{equation}
\lambda^{2} + \frac{\partial^{2} V}{\partial x^{2}}\big|_{x=x_{0}} = 0.
\label{eq:32}
\end{equation}

From Eq.(\ref{eq:32}) we obtain that if $\partial^{2} V / \partial x^{2}|_{x_{0}} < 0$, i.e., a diagram of a potential function is upper convex, the eigenvalues are real of opposite signs which corresponds to saddle points. In the opposite case, $\partial^{2} V / \partial x^{2}|_{x_{0}} > 0$, eigenvalues are purely complex and conjugate which correspond to centers. The saddle points in contrast to centers are structurally unstable critical points. While the saddles appear at maxima of $V(x)$, the centers correspond to minima. Therefore the diagram of the potential function with two maxima has to possess a minimum which corresponds to a center located between two saddles in the phase space. But the presence of the center in the phase space make the system structurally unstable. Hence the situation in which only one static critical point exists in finite domain is representing generic situation. The rest-exceptional cases are mere complicated and numerous and they interrupt the discussion of typical garden variety of cosmological dynamical systems.

Early history of investigations in the theory of dynamical systems has been dominated by searching for generic properties. They define a class of phase portraits that are far simpler that arbitrary one \cite{Peixoto:1962}. The part of the Kupka-Smale theorem states the genericity of G1 for critical points. A dynamical system has the property of G1 if all of its critical points are hyperbolic (or elementary). The center is an example of a non-hyperbolic critical point. Therefore the corresponding dynamical systems are exceptional or non-generic,

From eq. (\ref{eq:26}) we obtain that if $V(x)$ is an increasing function of its argument then in the interval of $x$ a universe is decelerating, while in the interval of $x$ in which $V(x)$ is a decreasing function of $x$, a universe is accelerating. At the maximum the universe starts accelerating which corresponds to redshift $z_{T} \colon 1 + z_{T} = x_{0}^{-1}$. This type of dynamical behavior in the neighborhood of redshift transition is a prototype of evolutional scenario of the dark energy epoch.

\section{Three generic scenarios of cosmological dynamics with dark energy}
\label{sec:5}

\begin{figure}[t]
\begin{center}
\includegraphics{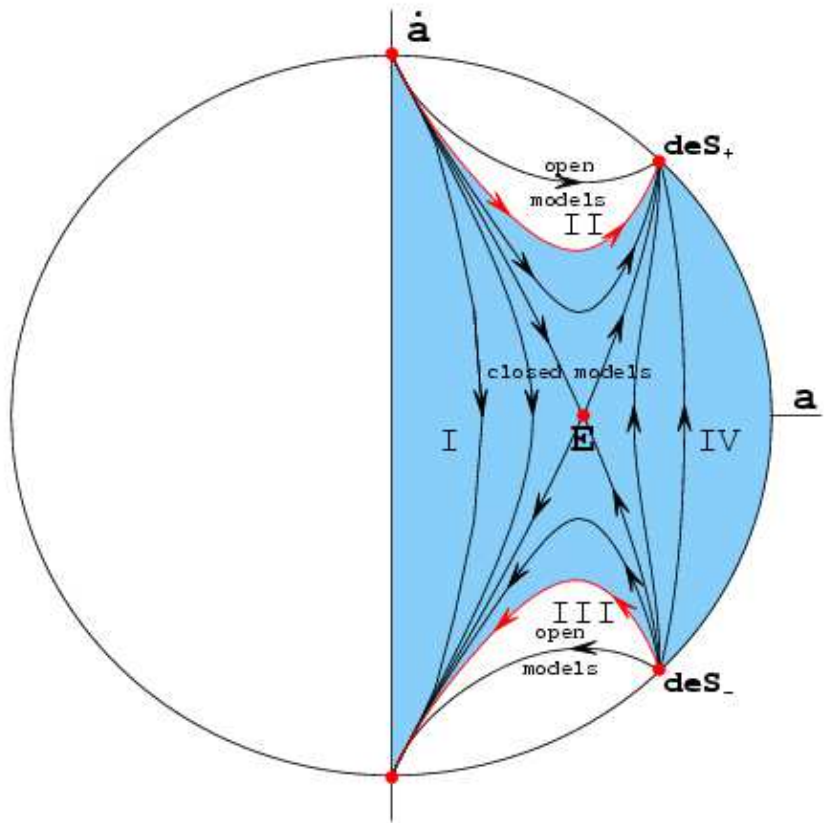}
\end{center}
\caption{The phase space $(a,\dot{a})$ portrait for the FRW model with dust and dark energy. We assume the existence of the moment during the evolution at which $\rho + 3p > 0$ is violated, where $\rho$ and $p$ are total energy and pressure. We have the static critical point marked as $E$ at finite domain and the critical points at infinity. The system is symmetric under the reflection $y \to -y$. The trajectory of the flat model divides the phase plane on two disjoint domains occupied by closed and open models. Points (always static) at finite domain remains intersection of boundary of the strong energy condition and $a$--axis. The points at infinity are located on the trajectory of the flat model and circle at infinity. For large $a$ the potential function looks like in the model with the cosmological constant $V \propto a^{2}$. The deSitter model deS$_{+}$ is the global attractor in the future. The trajectories coming to this point are perpendicular to the circle at infinity $S^{1}$. The model is structurally stable because contains all structurally stable critical points (non-hyperbolic).}
\label{fig:1}
\end{figure}

\begin{figure}[t]
\begin{center}
\includegraphics{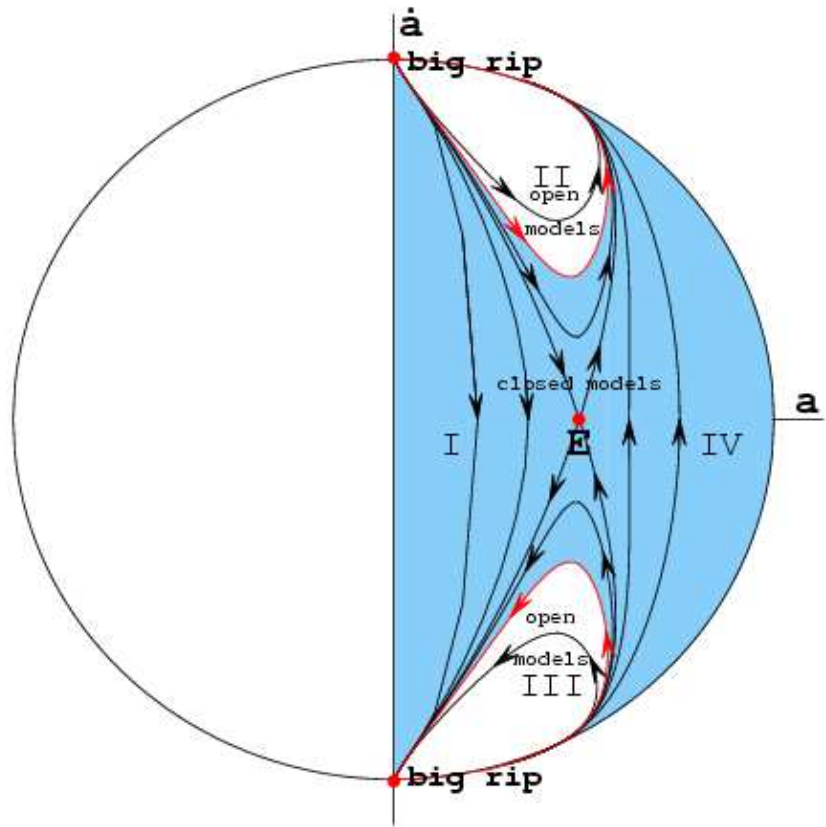}
\end{center}
\caption{The phase portrait for (\ref{eq:8}) (see description of Fig.~\ref{fig:1}) in the case of $V \propto a^{m}$ and $m > 2$ for large $a$. This is a representative and structurally stable model. The trajectories are tangent to the circle at infinity ($\ud \dot{a}/\ud a = 0$ as $a \to \infty$).}
\label{fig:2}
\end{figure}

\begin{figure}[t]
\begin{center}
\includegraphics{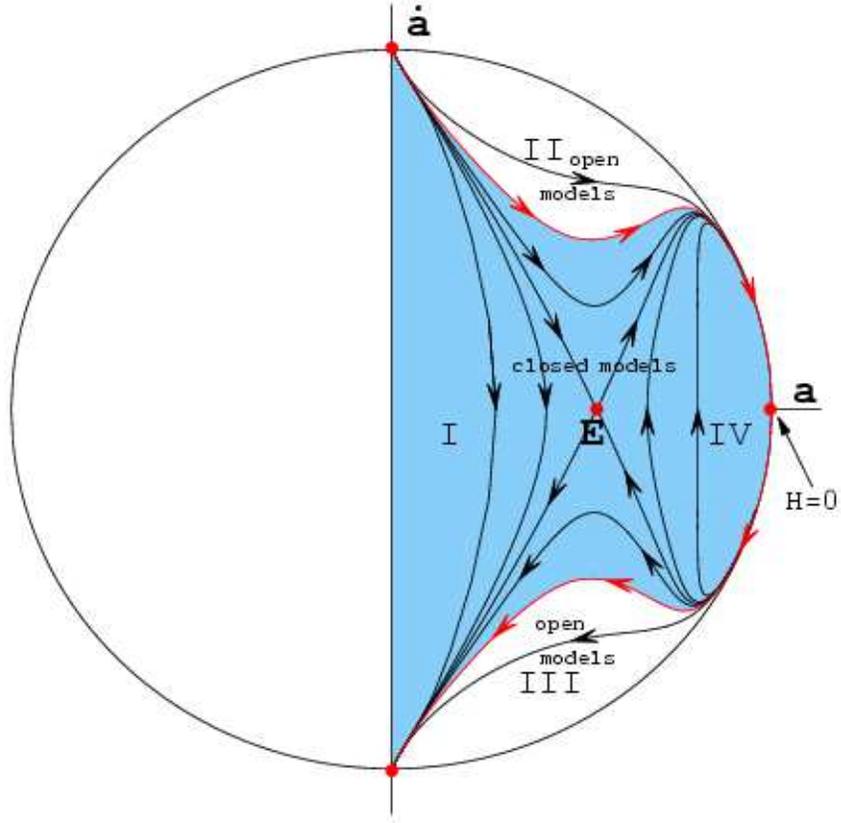}
\end{center}
\caption{The phase portrait for the model under consideration (for assumption see the description of Fig.~\ref{fig:1}) in the case of $V(a) \propto a^{m}$ and $m < 2$ how $a \to \infty$. In this case instead of the deS attractor the degenerated critical point $a = \infty$, $\dot{a} = 0$ ($H=0$) is present on the circle. This system is structurally unstable.}
\label{fig:3}
\end{figure}

The results of the qualitative analysis of the dynamical system are summarized and presented in Fig.~\ref{fig:1}--\ref{fig:3}. While all phase portraits are topologically equivalent (modulo homeomorphism preserving an orientation of the phase curve) at the finite domain, some differences are in the location of the critical points at the circle at infinity. There are only three prototypes of dynamical behavior of the model.

On the phase portraits we consider the plane $(a,\dot{a})$ where the coordinate $a$ is positive. All these solutions which correspond to the flat universe are located on the curve $\dot{a}^{2} = - 2V(a)$. Trajectories going inside of two branches of this curve correspond to closed models while those moving outside correspond to open models. 

One can distinguish the vertical boundary line $\{ a \colon \partial V/\partial a = 0\}$ which separates the solutions with acceleration from those without acceleration. The intersection of this boundary with the $a$--axis is at the saddle which is quite similar to Einstein's static universe. The cosmological singularities lie on the circle at infinity $a^{2} + \dot{a}^{2} = \infty$. They always appear as a intersection of a trajectory of the flat model and the circle at infinity. On the circle at infinity $S^{1}$ which compactifies the plane $\mathbf{R}^{2}$ to the projective plane, the effects of curvature are negligible.

Let us concentrate on the behavior of trajectories at the finite domain which is identical for all cases in Fig.~\ref{fig:1}--\ref{fig:3}. We find only one static critical point $a_{0}$ located on the $a$-axis which corresponds to a maximum of the potential function. Trajectories moving in the region $a < a_{0}$ confined by separatrices approaching and escaping saddle (on the left from saddle point denoted as $I$) correspond to the closed universes expanding to the maximal size and then contracting to the final singularity. Note that they correspond to the closed universes which never undergo an acceleration phase during the whole evolution. There exists also two another types of evolutional paths for closed models. They are (we describe expanding trajectories) located between the upper branch of flat trajectory and the separatrices which lies in the region $y > 0$. They start from the standard singularity $a=0$, $\dot{a} = \infty$ then reach a point of maximal expansion $\dot{a}$ and continue expansion to the deS$_{+}$ node (or state $H=0$ in the case in Fig.~\ref{fig:3}).

Another type of evolution experienced by closed models lies in the region $(IV)$ which boundaries form two separatrices, one coming from the saddle point to deS$_{+}$ and other one coming out the singularity and approaching the saddle. The trajectories located in this region describe the evolution from the state $a=\infty$, $\dot{a}=\infty$ ($\dot{a}=0$ in the exceptional case in Fig.~\ref{fig:3}) deS$_{-}$, i.e., the contracting deSitter model; next they reach a point of minimal contraction $\dot{a}=0$ and then they expand to deS$_{+}$ model if $V(a) \propto a^{m}$ and $m=2$ or reach a state $\dot{a}=\infty$, $a=\infty$ if $m > 2$. In other words the trajectories running in the region ($IV$) undergo a contraction from the unstable contracting deSitter node toward the stable deS$_{+}$ model -- the expanding deSitter model only if $m=2$ (i.e., the cosmological constant case $H=$ const). In the other case there is no deS$_{+}$ as a global attractor in the future ($H=\infty$). It should be pointed out that the existence of deS$_{+}$ attractor in the future is the unique property of the dark energy model for which $V \propto \rho_{x} x^{2}/6$ and $\rho_{x}(\infty) = $ const $ > 0$. If $m < 2$ and $V \propto x^{m}$ as $x$ goes to infinity then trajectories start and land at the degenerated critical point ($x = \infty$, $\dot{x} = 0$) which is of course the non-generic critical point. It is interesting that the location as well as the type of the critical points at infinity can be determined from asymptotic behavior of the potential function as $x \to \infty$ (or $x \to 0$).

Let us briefly comment the generic case when $V(x) \propto x^{2+\varepsilon}$, $\varepsilon > 0$. Then in the future evolution of the model with $\varepsilon > 0$ instead of deS phase there is unwanted type of a singularity, so called a big-rip singularity. Because this critical point is a stable node big-rip singularities are generic. Note that this type of final behavior is a global attractor for all open, closed and flat models. All these models undergo the transition from the decelerating to accelerating phases. All these models also undergo the so-called loitering phase like for the Lemaitre model \cite{Sahni:1992}. In the language of the potential function if the particle universe starts off with just enough kinetic energy to take it up to the top of the potential then value of kinetic energy close to this critical value gives rise to the loitering phase. Finally the existence of a loitering phase is a generic property of the models with dynamical dark energy.

The phase of possible phase portraits which are generic can be classified by the construction of representative cases (Fig.~\ref{fig:1} and \ref{fig:2}) which are non equivalent topologically.

\section{Conclusions}

The main message of this paper was to note that there exists a systematic methods of classification and investigations of the FRW models with the dynamical equation of state in the quite general form $p = w(a) \rho$. It was demonstrated that such class of models can by reduced to a certain two dimensional dynamical system. One of the features of such reduction is the possibility of representation dynamics as a motion of a fictitious particle--universe in one-dimensional potential. The method of Hamiltonian dynamical systems can serve as a natural tool for investigation of all solutions for all possible initial conditions. Then one is able to see how large is the class of solutions (labelled by the inset of initial conditions) leading to the desired property. We consider how large is the set of initial conditions which gives rise to the big-rip singularity.

In the applications of dynamics in various fields of science, the dynamics---that is dynamical system---can never by specified exactly. The dynamical system as a model of dynamics might be useful any way, if it can describe features of the phase portrait that persist when the vector field is allowed to move around. This idea emerged in the early history of modern dynamical system theory is called structural stability \cite{Smale:1980}.

The idea of structural stability originated with Andronov and Pontryagin in 1937. Their work on planar systems was extended with the Peixoto theorem which completely characterizes the structurally stable systems on a compact, two dimensional manifold and establishes that they are generic. The dynamical system is said to be structurally stable if close dynamical systems (in some metric sense) are topologically equivalent to it. In the case of two-dimensional dynamical systems on the Poincar{\'e} sphere there is an easy test for the structural stability of the global phase portrait of planar polynomial system. In particular, the phase portraits on the Poincar{\'e} sphere are structurally unstable if there are non-hyperbolic points at infinity or if there is a trajectory connecting a saddle on the equator if the Poincar{\'e} sphere to another saddle on $S^{2}$ (\cite{Perko:1991} p.322).

If we assume the existence of the moment during the evolution of the universe at which the strong energy condition is violated, then structural stability of the system requires the presence of saddles only (in the general case limit cycles and nodes are admissible). As a result we obtain three distinguished generic cases. The structural stability at infinity distinguishes only two phase portraits. It becomes in good agreements with phase portraits obtained from the reconstructed potential by using SN Ia \cite{Szydlowski:2004,Szydlowski:2004b}.

\end{document}